\documentclass[12pt]{iopart}
\usepackage{iopams}  

\def\oint{{\relax \int\kern -1. em O}}

\def\ba{\begin{eqnarray}}
\def\ea{\end{eqnarray}}
\newcommand{\eq}{\begin{equation}}
\newcommand{\feq}{\begin{equation}}

\def\Z{\mathbb Z}

\newcommand{\bea}{\begin{eqnarray}}
\newcommand{\eea}{\end{eqnarray}}

\font\tengoth=eufm10 \font\sevengoth=eufm7 \font\fivegoth=eufm5
\newfam\gothfam
\textfont\gothfam=\tengoth \scriptfont\gothfam=\sevengoth
  \scriptscriptfont\gothfam=\fivegoth

\def\<#1>{\langle#1\rangle}

\def\til{\widetilde}

\def\Z{\mathbb Z}

\def\summn{\frac{\sum_{i=1}^n m_i}{n}}

\def\suc{\sum_{i=1}^n c_i}
\def\sumc{c_0+\sum_{i=1}^n m_i c_i}

\def\Dtil{D_0+\sum_{i=2}^n D_i (m_i-m_1)}

\newenvironment{tabla}{\fontsize{8}{17}\selectfont}{}
\def\bt{\begin{tabla}}
\def\et{\end{tabla}}

\begin{document}
\jl{1}

\title[Shape Invariant potentials 
depending on $n$ parameters]{Shape Invariant potentials 
depending on $n$ parameters transformed by 
translation\footnote[1]{To appear in J. Phys. A: Math. Gen. (2000)}}

\author{Jos\'e F. Cari\~nena\footnote[2]{E-mail:\ {\tt jfc@posta.unizar.es}} and Arturo Ramos\footnote[3]{E-mail:\ {\tt arrg@posta.unizar.es}}}

\address{Departamento de F\'{\i}sica Te\'orica. Facultad de Ciencias. \\
Universidad de Zaragoza, 50009, Zaragoza, Spain.}

\begin{abstract}
Shape Invariant potentials in the sense of [Gendenshte\"{\i}n L.\'E., 
JETP Lett. {\bf 38}, (1983) 356] which depend on more than two
parameters are not know to date. In [Cooper F., Ginocchio J.N. and Khare A.,
Phys. Rev. {\bf 36 D}, (1987) 2458] was posed the problem of finding
a class of Shape Invariant potentials which depend on $n$ parameters
transformed by translation, but it was not solved. We analyze 
the problem using some properties of the Riccati equation and
we find the general solution.
\end{abstract}




\section{Introduction}

There has been much interest in the search of exactly solvable problems
in Quantum Mechanics from the early days of the theory to date. To this respect,
the Factorization Method introduced by Schr\"odinger \cite{Sch1,Sch2,Sch3} and 
later developed by Infeld and Hull \cite{InfHul} has been shown to be very
efficient. Later, the introduction of Supersymmetric Quantum Mechanics by 
Witten \cite{Witten81} and the concept of Shape Invariance by 
Gendenshte\"{\i}n \cite{Gen} have renewed to great extent the interest in the
subject. For an excellent review, see \cite{CoopKhaSuk}.

In particular, Shape Invariant problems have been shown to be exactly solvable,
and it was observed that a number of known exactly solvable potentials belonged 
to such a class. The natural question which arose was whether all exactly
solvable problems have the property of being Shape Invariant in the 
sense of \cite{Gen}. This question has been treated in an interesting paper
several years ago \cite{CoopGinKha}. There, the Natanzon class of 
potentials \cite{Nat} was investigated in detail. Following that line of 
reasoning, the authors gave a classification of Shape Invariant potentials
whose parameters are transformed by translation. They proposed the general
case which depends on an arbitrary but finite number $n$ of parameters,
and established the equations to be solved in order to find such a class.
But they asserted to have failed to find any solution of the equations.

For several years this class of Shape Invariant potentials has been 
considered to be a good candidate to enlarge the class of known solutions
of the Shape Invariance condition, see e.g. \cite{BarMax,BarDutGanKhaPagSuk}.
But the solutions are not known so far. 

As it seems to be an interesting problem 
we have analyzed it carefully and we have proved
that it is possible to find in an easy way the solution.
The main point is to use in an appropriate way some interesting
properties of a related Riccati equation. 
As a consequence, the aim of this paper is to answer to the question proposed 
in~\cite{CoopGinKha}. 

The organization of the paper is as follows.
After a quick description of the problem of Shape Invariance in Section 2,  
we will develop in Section 3
the mathematical study of a particularly interesting first order ordinary 
differential equation system of key importance for the problem. 
Then we will proceed to study in Section 4 
the problem of Shape Invariant potentials depending on $n$ parameters.
We will do some {\emph{Ans\"atze}} for the superpotentials assuming 
translations as the transformation law for the parameters, including the
one proposed in \cite{CoopGinKha} and its more immediate generalizations.
The results are presented in some tables.

\section{Shape in\-va\-rian\-ce and the Fac\-tori\-zation Me\-thod}

We recall some basic ideas of the theory of related operators, the concept
of partner potentials and Shape Invariance.
Two Hamiltonians 
\begin{equation}
H=-\frac{d^2}{dx^2}+V(x)\,,\ \ \ \til H=-\frac{d^2}{dx^2}+\til V(x)\,,
\label{defHHtil}
\end{equation}
are said to be related whether there exists an operator $A$ such 
that $AH=\til H A$, where $A$ need not to be invertible.
If we  assume that 
\begin{equation}
A=\frac{d}{dx}+W(x)\,,
\label{defA}
\end{equation}
then, the relation $AH=\til H A$
leads to
\begin{equation}
W(V-\til V)=-W''-V'\,,\quad
V-\til V=-2 W'\,,             \label{1stset}
\end{equation}
while the relation $HA^{\dagger}=A^{\dagger}\til H$
leads to
\begin{equation}
W(V-\til V)=W''-\til V'\,,    \quad
V-\til V=-2 W'\,.             \label{2ndset}
\end{equation}
One can easily integrate both pair of equations; we obtain 
$$
V=W^2-W'+c\,,\quad
\til V=W^2+W'+d\,,
$$
where $c$ and $d$ are constants. But taking into account the equation
$V-\til V=-2 W'$ we have $c=d$. Therefore (see e.g \cite{CarMarPerRan}),  
 two Hamiltonians $H$ and $\til H$ of the form
 (\ref{defHHtil}) can be related by a first order
differential operator $A$ like (\ref{defA}) if and only if
there exists a real constant $d$ such that $W$ satisfies the
pair of Riccati equations
\begin{equation}
V-d=W^2-W'\,,    \quad  \til V-d=W^2+W'\ , \label {ricV}
\end{equation}
and then the  Hamiltonians can be factorized
as
\begin{equation}
H=A^\dagger A+d\,,\ \ \ \til H=A A^\dagger+d\,.
\end{equation} 
Using  equations in (\ref{ricV}) 
we obtain the equivalent pair
\begin{equation}
\til V-d=-(V-d)+2W^2\,,       \quad \til V=V+2 W'\,.    \label{relVVtilder} 
\end{equation}
The potentials $\til V$ and $V$ are usually  said to be
\emph{partners}.

We would like to remark that these equations have an intimate relation with
what it is currently known as \emph{Darboux transformations} in the context
of one dimensional or Supersymmetric Quantum Mechanics. In fact, it is easy
to prove that the first of the equations (\ref{ricV}) can be transformed into 
a Schr\"odinger equation $-\phi^{\prime\prime}+(V(x)-d)\phi=0$ by means of the 
change $-\phi^{\prime}/\phi=W$, and by means of 
$\til\phi^{\prime}/\til\phi=W$ the second of (\ref{ricV}) transforms into
$-\til\phi^{\prime\prime}+(\til V(x)-d)\til\phi=0$. The relation between
$V$ and $\til V$ is given by (\ref{relVVtilder}).
Obviously, $\phi \til\phi=1$, up to a non--vanishing constant factor. 
It is also worth noting that these Schr\"odinger equations 
express that $\phi$ and $\til\phi$ are respective eigenfunctions 
of the Hamiltonians (\ref{defHHtil}) for the eigenvalue $d$. 
These are the essential points of the mentioned Darboux transformations, 
as exposed e.g. in \cite[pp. 7,\,24]{MatSal91}.

The concept of \emph{Shape invariance} introduced by 
Gendenshte\"{\i}n \cite{Gen}: $V$  is assumed to depend on certain
set of parameters and equations (\ref{ricV}) 
define $V$ and $\til V$  in terms of a superpotential $W$.
The  condition for a partner  $\til V$ to be of the same form as $V$ 
but for a different choice  of the values of the parameters involved in $V$, 
is called Shape Invariance condition \cite{Gen}.

More explicitly, if
$V=V(x,a)$ and $\til V=\til V(x,a)$, where $a$ denotes a set of parameters,
Gendenshte\"{\i}n \cite{Gen} showed that if we assume the further relation
between $V(x,a)$ and $\til V(x,a)$ given by
\begin{equation}
\til V(x,a)=V(x,f(a))+R(f(a))\,,
\label{SIGed}
\end{equation}
where $f$ is a transformation 
of the set of parameters $a$ and $R(f(a))$ is a remainder 
not depending on $x$, then the complete spectra of
the Hamiltonians $H$ and $\til H$ can be found easily.
 Just writing the $a$--dependence
the equations (\ref{ricV}) become  
\begin{equation}
V(x,a)-d=W^2-W'\,,           \quad
\til V(x,a)-d=W^2+W'\,.       \label{ricVSI}  
\end{equation}

Therefore, we will assume that $V(x,a)$ and $\til V(x,a)$ are 
obtained from a superpotential function $W(x,a)$ by means of
\begin{equation}
\!V(x,a)\!-\!d\!=\!W^2(x,a)\!-\!W'(x,a)\,,                   \quad
\til V(x,a)\!\!-d\!=\!W^2(x,a)\!+\!W'(x,a)\,.  \label{ricVSIsp}        
\end{equation}  
The  Shape Invariance property   in the sense of
\cite{Gen}  requires the further condition (\ref{SIGed}) to be
satisfied.

The relationship of a slight generalization of the 
Factorization Method developed by Infeld and Hull \cite{InfHul} with 
the Shape Invariance theory has been explicitly established in \cite{CarRamdos}.
There, the following identifications between the symbols used in the
Factorization Method 
and those of Shape Invariance problems were found: 
\ba
\til V(x,a)-d&=&-r(x,f(a))-L(a)\,,              \label{idVtilrL_gen}    \\
V(x,a)-d&=&-r(x,a)-L(a)\,,                      \label{idVrL_gen}       \\ 
W(x,a)&=&k(x,a)\,,                              \label{idWk_gen}        \\
R(f(a))&=&L(f(a))-L(a)\,.                       \label{idRL_gen}
\ea

\section{General solution of equations \lowercase{$y^2+y'=a$, 
$zy+z'=b$}\label{solgyz}}

We will study next the general solution of a certain first order 
ordinary differential equation system. It  will play a key role 
in the derivation of the main subject in this paper. The system is
\ba
y^2+y'&=&a\,,                                   \label{ric_y}\\
y z+z'&=&b\,,                                   \label{lin_z}
\ea 
where $a$ and $b$ are real constants and the prime denotes derivative
respect to $x$. The equation (\ref{ric_y}) is a Riccati equation with 
constant coefficients, meanwhile (\ref{lin_z}) is an  inhomogeneous
linear first order differential equation for $z$, provided the function 
$y$ is known. 
The  general solution of (\ref{lin_z}) is 
easily obtained  once we know the solutions of (\ref{ric_y}), e.g. by means of 
\begin{equation}
z(x)=\frac{b\,\int^x\exp\big\{\int^\xi y(\eta)\,d\eta\big\}\,d\xi+D}
{\exp\big\{\int^x y(\xi)\,d\xi \big\}}\,,       \label{sol_z}
\end{equation}
where $D$ is an integration constant \cite{CarRamdos}.
 
The general Riccati equation  
\begin{equation}
\frac{dy}{dx}=a_2(x) y^2+a_1(x) y+a_0(x)\,,  \label{ric_y_gen}
\end{equation}
where $a_2(x)$, $a_1(x)$ and $a_0(x)$ are differentiable functions of 
the independent variable $x$, has very interesting properties. 
It is to be remarked that in the most general case there is no way of 
writing the general solution by using some quadratures, 
but  one can integrate it completely
 if one particular solution $y_1(x)$ of (\ref{ric_y_gen}) is 
known. Then,
 the change of variable (see e.g. \cite{Dav,Mur})
\begin{equation}
u=\frac{1}{y_1-y}\,,\quad\mbox{with inverse} 
\quad y=y_1-\frac{1}{u}\,,                      \label{ch_1sol_usual}
\end{equation}
transforms (\ref{ric_y_gen}) into the inhomogeneous first order linear
equation
\begin{equation}
\frac{du}{dx}=-(2\,a_2\,y_1+a_1)u+a_2\,,
\end{equation}
which can be integrated by two quadratures. An alternative
change of variable was also proposed recently \cite{CarRam}:
\begin{equation}
u=\frac{y\,y_1}{y_1-y}\,,\quad\mbox{with inverse} 
\quad y=\frac{u\,y_1}{u+y_1}\,.   \label{ch_1sol_nues}
\end{equation}
This change transforms (\ref{ric_y_gen}) into the 
inhomogeneous first order linear equation
\begin{equation}
\frac{du}{dx}=\bigg(\frac{2\,a_0}{y_1}+a_1\bigg)u+a_0\,,
    \label{lin_ch_1sol_nues}
\end{equation}
which is integrable by two quadratures, as well. 
 We also remark that the general 
Riccati equation (\ref{ric_y_gen}) admits the identically vanishing 
function as a solution if and only if $a_0(x)=0$ for all $x$ in the domain
of the solution. 

But the most important  property of Riccati equation is that 
when  three particular solutions of (\ref{ric_y_gen}), 
$y_1(x),\,y_2(x),\,y_3(x)$ are known, the general solution $y$ can be
automatically 
written, by means of the formula
\begin{equation}
y=\frac {y_2(y_3-y_1)\,k+y_1(y_2-y_3)}{(y_3-y_1)\,k+y_2-y_3}\ ,
                                                \label{solv_y}
\end{equation}
where $k$ is a constant determining each solution. 
As an example, it is easy to check that
$y|_{k=0}=y_1$, $y|_{k=1}=y_3$ and that the solution $y_2$ is obtained as
 the limit of $k$ going to  $\infty$.
For more information on geometric and group theoretic aspects of
 Riccati equation see e.g. \cite{CarMarNas,CarRam,CarGraRam,PW1}.

We are interested here in the simpler case of the
Riccati equation with constant coefficients (\ref{ric_y}). 
The general equation of this type is
\begin{equation}
\frac{dy}{dx}=a_2 y^2+a_1 y+a_0\,,  \label{ric_y_const}
\end{equation}
where $a_2$, $a_1$ and $a_0$ are now real constants, $a_2\neq 0$. 
This equation, unlike the general Riccati equation
 (\ref{ric_y_gen}),
is always integrable by quadratures, and the form of the 
solutions depends strongly on the sign of the discriminant 
$\Delta=a_1^2-4a_0 a_2$. This can be seen by separating the differential
equation (\ref{ric_y_const}) in the form
$$
\frac{dy}{a_2 y^2+a_1 y+a_0}=\frac{dy}
{a_2\bigg(\big(y+\frac{a_1}{2\,a_2}\big)^2
-\frac{\Delta}{4\,a_2^2}\bigg)}=dx\,.
$$
Integrating (\ref{ric_y_const}) in this way we obtain non--constant 
solutions. 

Looking for constant solutions of (\ref{ric_y_const}) amounts to solve
an algebraic second degree equation. So, if $\Delta>0$ there will be
two different real constant solutions, when  $\Delta=0$ there is
only one constant real solution and if $\Delta<0$ we have no constant
real solutions at all.

These properties may be used  for  finding the general
solution of (\ref{ric_y}). For this equation 
the discriminant $\Delta$ is just $4a$. 
Then,  
if $a>0$ we can write $a=c^2$, where $c>0$ is a real number. The 
non--constant particular solution
\begin{equation}
y_1(x)=c\tanh(c(x-A))\,,                \label{sp_y1_a>0}
\end{equation}
where $A$ is an arbitrary integration constant, 
is readily found by direct integration. In addition, there exist two
different constant real solutions,
\begin{equation}
y_2(x)=c\,,\quad\quad y_3(x)=-c\,.      \label{scp_y1_a>0}
\end{equation}  
The general solution obtained 
using formula (\ref{solv_y}), is
\begin{equation}
y(x)=c\,\frac{B\,\sinh(c(x-A))-\cosh(c(x-A))}
{B\,\cosh(c(x-A))-\sinh(c(x-A))}\,,
                                                        \label{gen_y_a>0}
\end{equation}
where $B=(2-k)/k$, $k$ being the arbitrary constant 
in (\ref{solv_y}). Substituting in (\ref{sol_z})
we obtain the general solution for $z(x)$,
\ba
z(x)=\frac{\frac{b}{c}\{B\,\sinh(c(x-A))-\cosh(c(x-A))\}+D}
{B\,\cosh(c(x-A))-\sinh(c(x-A))}\,,
\ea 
where $D$ is a new integration constant. 

For the case  $a=0$, a particular solution is
\begin{equation}
y_1(x)=\frac 1{x-A}\,,                          \label{spar_a=0}
\end{equation}
where $A$ is an integration constant. 
If we apply  the change 
of variable (\ref{ch_1sol_nues}) with $y_1$ given by (\ref{spar_a=0}),
then  (\ref{ric_y}) with $a=0$ transforms into 
$du/dx=0$. 
Then, the general solution for (\ref{ric_y}) with $a=0$ is
\begin{equation}
y(x)=\frac{B}{1+B(x-A)}\,,
                                                        \label{gen_y_a=0}
\end{equation}
with $A$ and $B$ being arbitrary integration constants. 
Substituting in (\ref{sol_z})
we obtain the general solution for $z(x)$ in this case,
\ba
z(x)=\frac{b(\frac B 2 (x-A)^2+x-A)+D}{1+B(x-A)}\,,
\ea 
where $D$ is a new integration constant. 

If now $a=-c^2<0$,
where $c>0$ is a real number we  find by direct integration the 
particular solution 
\begin{equation}
y_1(x)=-c\tan(c(x-A))\,,                \label{spar_a<0}
\end{equation}
where $A$ is an arbitrary integration constant.
With either the change of variable
(\ref{ch_1sol_usual}) or alternatively (\ref{ch_1sol_nues}), with $y_1(x)$
given by (\ref{spar_a<0}) we get  the general solution of 
(\ref{ric_y}) for $a>0$
\begin{equation}
y(x)=-c\,\frac{B\,\sin(c(x-A))+\cos(c(x-A))}
{B\,\cos(c(x-A))-\sin(c(x-A))}\,,
                                                        \label{gen_y_a<0}
\end{equation}
where $B=cF$, $F$ arbitrary constant. 
Substituting in (\ref{sol_z})
we obtain the general solution for $z(x)$ in this case,
\ba
z(x)=\frac{\frac{b}{c}\{B\,\sin(c(x-A))+\cos(c(x-A))\}+D}
{B\,\cos(c(x-A))-\sin(c(x-A))}\,,
\ea 
where $D$ is a new integration constant. 

These solutions can be written in many mathematically equivalent ways. 
We have tried to give their simplest form and in 
such a way that the symmetry between the solutions for the case $a>0$ and
$a<0$ were clearly recognized. Indeed, the general solution of (\ref{ric_y}) 
for $a>0$ can be transformed into that of the case $a<0$ by means of
the formal changes $c\rightarrow ic$, $B\rightarrow iB$ and the identities
$\sinh(ix)=i\sin(x)$, $\cosh(ix)=\cos(x)$. 
The results are summarized in Table~\ref{sols_gens}.

Looking at  the 
general solution of (\ref{ric_y}) for $a>0$, i.e. equation
 (\ref{gen_y_a>0}),
one could be tempted to write it in the form of a logarithmic derivative,
$$
y(x)=\frac{d}{dx}\log|B\,\cosh(c(x-A))-\sinh(c(x-A))|\,.
$$
This is  equivalent except for $B\rightarrow\infty$. 
In fact, if we want to calculate
$$
\lim_{B\to\infty}\frac{d}{dx}\log|B\,\cosh(c(x-A))-\sinh(c(x-A))|
$$
we \emph{cannot} interchange the limit with the derivative,
 otherwise we would get a wrong result. 
But this limit for $B$ is particularly important since when taking
it in (\ref{gen_y_a>0}), we recover the particular solution (\ref{sp_y1_a>0}).
A similar thing happens in the general solutions (\ref{gen_y_a=0}) and
(\ref{gen_y_a<0}). When taking the limit $B\to\infty$ we recover, 
respectively, the particular solutions (\ref{spar_a=0}) and (\ref{spar_a<0}),
from which we have started.
Both of (\ref{gen_y_a=0}) and (\ref{gen_y_a<0})
can be written in the form of a logarithmic derivative, but then the limit
$B\to\infty$ could not be calculated properly.

\section{Shape Invariant potentials depending on an arbitrary
number of parameters transformed by translation}
                
We will try now to generalize the class of possible factorizations
considered in \cite{InfHul,CarRamdos}. We analyze the 
possibility of introducing superpotentials  depending on an arbitrary
but finite number of  parameters $n$ which transforms by translation.
This will give in turn the still unsolved problem proposed in~\cite{CoopGinKha}.
 
More explicitly,  suppose that within the parameter space some 
of them transform according to 
\begin{equation}
f(a_i)=a_i-\epsilon_i\,,\quad\forall\,i\in \Gamma\,,
\label{tas}
\end{equation}
and the remainder according to
\begin{equation}
f(a_j)=a_j+\epsilon_j\,,\quad\forall\,j\in \Gamma'\,,
\label{tas_+}
\end{equation}
where $\Gamma\cup\Gamma'=\{1,\,\dots,\,n\}$, 
and  $\epsilon_i\neq 0$ for all $i$. Using a reparametrization,
one can normalize each parameter in units of $\epsilon_i$, that is, 
we can introduce the new parameters
\begin{equation}
m_i=\frac{a_i}{\epsilon_i}\,,\quad\forall\,i\in \Gamma\,,
\quad \mbox{and}\quad m_j=-\frac{a_j}{\epsilon_j}\,,\quad\forall\,j\in \Gamma'\,,
\label{parmi}
\end{equation}
for which the transformation law reads, 
with a slight abuse of the notation $f$,
\begin{equation}
f(m_i)=m_i-1\,,\ \ \ \forall\,i=1,\,\dots,\,n\,.
\label{tms}
\end{equation}
Note that with these normalization, the initial values of each $m_i$ are
defined by some value in the interval $(0,1]\ \pmod{\Z}$.

We will use the notation $m-1$ for the $n$--tuple 
$m-1=(m_1-1,\,m_2-1,\,\dots,\,m_n-1)$. The transformation law 
for the parameters (\ref{tms}) is just 
a particular case of a more general  transformation considered in
\cite{CarRamdos}. As a corollary of a result proved there we have  
following one. 
\emph{The problem of finding the square integrable solutions of the
equation 
\begin{equation}
\frac{d^2 y}{dx^2}+r(x,m)y+\lambda y=0\,,
\label{SODE_gen}
\end{equation}
according to the generalization of the Infeld and Hull Factorization
Method treated in \cite[Sec. 3]{CarRamdos}, 
is equivalent to that of solving the discrete eigenvalue problem 
of Shape Invariant potentials
in the sense of \cite{Gen} depending on the same 
$n$--tuple of parameters $m\equiv (m_1,\,m_2,\,\dots,\,m_n)$ which transform
according to (\ref{tms})}.

\smallskip

In order  to find solutions for these problems, 
we should find solutions of the difference-differential equation
\begin{equation}
k^2(x,m+1)-k^2(x,m)
+\frac{dk(x,m+1)}{dx}+\frac{dk(x,m)}{dx}=L(m)-L(m+1)\,,
\label{lhs_np}
\end{equation}
where now $m=(m_1,\,m_2,\,\dots,\,m_n)$ 
denotes the set of parameters and $m+1$ means
$m+1=(m_1+1,\,m_2+1,\,\dots,\,m_n+1)$,
and $L(m)$ is some function to be determined, related to $R(m)$
by $R(m)=L(m)-L(m+1)$. Equation (\ref{lhs_np}) is essentially
equivalent to the Shape Invariance condition 
$\til V(x,m)=V(x,m-1)+R(m-1)$ for problems defined by (\ref{tms}) 
\cite{CarRamdos}. We would like to remark that (\ref{lhs_np}) always
has the trivial solution $k(x,m)=h(m)$, for every arbitrary function 
$h(m)$ of the parameters only. 

Our first assumption for the dependence of $k(x,m)$ on $x$ and $m$
will be a generalization of the one
used for the case of one parameter introduced in \cite{InfHul}, 
\begin{equation}
k(x,m)=k_0(x)+m\,k_1(x)\,,\label{mlin}
\end{equation}
where $k_0$ and $k_1$ are functions of $x$ only. 
The generalization to $n$ parameters is 
\begin{equation}
k(x,m)=g_0(x)+\sum_{i=1}^n m_i g_i(x)\,. \label{mlin_np}
\end{equation} 
This form for $k(x,m)$ is exactly the same as the one proposed
in \cite[Eqs. (6.24)]{CoopGinKha} taking into account 
(\ref{parmi}) and (\ref{tms}), up to a slightly different notation.
Substituting into (\ref{lhs_np}) we obtain
\ba
&&L(m)-L(m+1)                                   \nonumber\\                                      
&&=2\sum_{j=1}^n m_j\bigg(g'_j+g_j\sum_{i=1}^n g_i\bigg)
+\sum_{j=1}^n (g'_j+g_j\sum_{i=1}^n g_i)        
+2 \bigg(g'_0+g_0\sum_{i=1}^n g_i\bigg)\,.      \label{lhs_npar}
\ea
Since the coefficients of the powers of each $m_i$ have to 
be constant, we obtain the following first order 
differential equation system to be satisfied,
\ba
&&g'_j+g_j\sum_{i=1}^n g_i=c_j\,,\ \ \ \forall\,j\in\{1,\,\dots,\,n\}\,,
                                                \label{eqsgi}\\
&&g'_0+g_0\sum_{i=1}^n g_i=c_0\,,
                                                \label{eqsg0}
\ea
where $c_i$, $i\in \{0,\,1,\,\dots,\,n\}$ are real constants.

The solution of the system can be found by using
barycentric coordinates for the $g_i$'s, that is, the functions
which separate the unknowns $g_i$'s in their
mass--center coordinates and relative ones.
Hence, we will make the following change of variables and use the  notations
\ba
g_{cm}(x)&=&\frac 1 n \sum_{i=1}^n g_i(x)\,,\label{gcm}\\ 
v_j(x)&=&g_j(x)-g_{cm}(x)               
=\frac 1 n \bigg(n g_j(x)-\sum_{i=1}^n g_i(x)\bigg)\,,\label{vj}\\
c_{cm}&=&\frac 1 n \sum_{i=1}^n c_i\,,  \label{ccm}
\ea
where $j\in\{1,\,\dots,\,n\}$. 
Note that not  all of the functions $v_j$ are now linearly independent, 
but only $n-1$ since $\sum_{j=1}^n v_j=0$.

Taking the sum of equations (\ref{eqsgi}) we obtain that $n g_{cm}$
satisfies the Riccati equation with constant coefficients
$$
n g'_{cm}+(n g_{cm})^2=n c_{cm}\,.
$$
On the other hand, we will consider the independent functions 
$v_j(x)$, $j\in\{2,\,\dots,\,n\}$ to complete the system. Using
equations (\ref{vj}) and (\ref{eqsgi}) we find
\ba
v'_j&=&\frac 1 n (n g'_j-\sum_{i=1}^n g'_i)             \nonumber\\
&=&\frac 1 n (g'_j-g'_1+g'_j-g'_2+\dots+g'_j-g'_j+\dots+g'_j-g'_n)
                                                        \nonumber\\
&=&-v_j n g_{cm}+c_j-c_{cm}\,,                          \nonumber
\ea
and we will take the corresponding equations from $2$ to $n$. 
The system
of equations (\ref{eqsgi}) and (\ref{eqsg0}) is written in the new
coordinates as
\ba
&&n g'_{cm}+(n g_{cm})^2=n c_{cm}\,,    \label{ricgcm}\\
&&v'_j+v_j n g_{cm}=c_j-c_{cm}\,,\ \ \forall\,j\in\{2,\,\dots,\,n\}\,,
                                        \label{linvj}\\
&&g'_0+g_0 n g_{cm}=c_0\,,              \label{ling0}
\ea 
and therefore the motion of the center of mass is decoupled from the
 other coordinates.
But we already know the general solutions of equation (\ref{ricgcm}),
which is nothing but the equation (\ref{ric_y}) studied in the preceding 
section with the identification of  $y$ and 
$a$ with $n g_{cm}$ and $n c_{cm}$, respectively. 
Therefore the  possible solutions depend on the
sign of $n c_{cm}$, that is, on the sign of the sum $\suc$ 
of all the constants
appearing in equations (\ref{eqsgi}). Moreover, all  the remaining 
equations (\ref{linvj}) and (\ref{ling0}) are linear differential 
equations of the form (\ref{lin_z}), identifying $z$ as
$v_j$ or $g_0$, and the constant $b$ as $c_j-c_{cm}$ or $c_0$,
respectively. The general solution of these equations is readily found 
once $n g_{cm}$ is known, by means of the 
formula (\ref{sol_z}) adapted to each case.
As a result  the general solutions for the variables $n g_{cm}$, $v_j$ and 
$g_0$ are directly found by just 
looking at Table~\ref{sols_gens} and making the proper substitutions.
The results are shown in Table~\ref{ngcmvjg0}.

Once the solutions of equations (\ref{ricgcm}), 
(\ref{linvj}) and (\ref{ling0}) are known it is easy 
to find the expressions for $g_i(x)$ and $g_0(x)$ by reversing
the change defined by (\ref{gcm}) and (\ref{vj}). 
It is easy to prove that it is indeed invertible 
with inverse change given by
\ba
g_1(x)&=&g_{cm}(x)-\sum_{i=2}^n v_i(x)\,,               \label{g1}\\
g_j(x)&=&g_{cm}(x)+v_j(x)\,,\ \ \ \forall\,j\in\{2,\,\dots,\,n\}\,.
                                                        \label{gj2n}
\ea
For each of the three families of solutions shown in 
Table~\ref{ngcmvjg0}, one can quickly find the
corresponding functions $g_i(x)$, $g_0(x)$, and hence 
the function $k(x,m)$ according to (\ref{mlin_np}). 
The results are shown in Table~\ref{sols_k_np_lin}.

We can now calculate the corresponding Shape Invariant partner
potentials by means of the formulas (\ref{ricVSIsp}), (\ref{idWk_gen})
and (\ref{idRL_gen}) adapted to this case. The results are shown 
in Table~\ref{sols_sipot_np}.

Let us comment on  the solutions for the function $k(x,m)$
in Table~\ref{sols_k_np_lin} and for the Shape Invariant 
potentials in Table~\ref{sols_sipot_np} we 
have just found. It is remarkable that the constants $c_i$, $c_0$, of
equations (\ref{eqsgi}), (\ref{eqsg0}) appear always in the solutions 
by means of the combination $\sumc$. 
On the other hand, $\til D$ does not change under the transformation  
$m_i\rightarrow m_i-1$ since it depends only on differences of the 
$m_i$'s. As $D_0,\,D_2,\,\dots,\,D_n$ are arbitrary constants, 
$\til D=\Dtil$ can be regarded as an arbitrary constant as well.  
It is very easy to check that the functions $k(x,m)$
satisfy indeed (\ref{lhs_np}), just taking into account that 
$n c_{cm}=\sum_{i=1}^n c_i$
and that when $n c_{cm}=C^2$, $\sum_{i=1}^n c_i/C=C$, meanwhile
$\sum_{i=1}^n c_i/C=-C$ when $n c_{cm}=-C^2$. Obviously, for
the case $n c_{cm}=0$ we have $\suc=0$. 
As we have mentioned already,
(\ref{lhs_np}) is essentially equivalent to the Shape Invariance
condition $\til V(x,m)=V(x,m-1)+R(m-1)$, but this last can be checked
directly. 
In order to do it, it may be useful to recall several relations 
that the functions defined in Table~\ref{ngcmvjg0} satisfy.
When $n c_{cm}=C^2$ we have
\ba
f'_{+}=C(1-f_{+}^2)=C(B^2-1)h_{+}^2\,,\quad\quad      
h'_{+}=-C f_{+} h_{+}\,,                      \nonumber
\ea
when $n c_{cm}=0$, 
\ba
f'_{0}=-B\,f_{0}^2\,,\quad\quad
h'_{0}=-B\,f_{0} h_{0}+1\,,                   \nonumber
\ea 
and finally when $n c_{cm}=-C^2$,
\ba
f'_{-}=C(1+f_{-}^2)=C(B^2+1)h_{-}^2\,,\quad\quad
h'_{-}=C f_{-} h_{-}\,,                       \nonumber
\ea
where the prime means derivative respect to $x$.
The arguments of the functions are the same as in the 
mentioned table and have been dropped out for simplicity.
 
When we have only one parameter, that is, $n=1$, one 
recovers the solutions for $k(x,m)=k_0(x)+m k_1(x)$ shown in the first 
column of \cite[Table~6]{CarRamdos}, and the corresponding 
Shape Invariant partner potentials of Table~7 in the same reference.

For all cases in Table~\ref{sols_sipot_np},
the formal expression of $R(m)$ is exactly the same,
but either $\suc=n c_{cm}$ have different sign or vanish.
Let us consider now the problem of how to
determine $L(m)$ from $R(m)$. The method 
does not provide the expression of $L(m)$ but of $L(m)-L(m+1)$. 
In fact, there is a freedom in determining this  function  $L(m)$.
Fortunately, for the purposes of Quantum Mechanics the relevant 
function is $R(m)$, from which the energy spectrum of Shape Invariant 
potentials in the sense of \cite{Gen} is calculated \cite{CarRamdos}.

However, let us show how this underdetermination appears. 
Since 
\ba
R(m)=L(m)-L(m+1)=2(\sumc)+\suc\, \label{RLsum}
\ea
is a polynomial in the $n$ parameters $m_i$, and we have considered
only polynomial functions of these quantities so far, $L(m)$ should be also a 
polynomial. It is of degree two, otherwise a simple calculation would show
that the coefficients of terms of degree 3 or higher must vanish. 
So, we propose 
$L(m)=\sum_{i,j=1}^n r_{ij}m_i m_j+\sum_{i=1}^n s_i m_i+t$, where 
$r_{ij}$ is symmetric, $r_{ij}=r_{ji}$. Therefore,
there are $\frac 1 2 n(n+1)+n+1$ constants to be determined. 
Then, making use of the symmetry of $r_{ij}$ in its indices we obtain
\ba
L(m)-L(m+1)
&=&-2\sum_{i,j=1}^n r_{ij} m_i-\sum_{i,j=1}^n r_{ij}-\sum_{i=1}^n s_{i}\ . 
                                                \nonumber
\ea
Comparing with (\ref{RLsum}) we find the following conditions to be satisfied 
$$
-\sum_{j=1}^n r_{ij}=c_i\,,
\ \ \ \forall\,i\in\{1,\,\dots,\,n\}\,,\ \ \ \mbox{and}
\ \ \ -\sum_{i=1}^n s_i=2c_0\,.                                         
$$
The first of these equations expresses the problem of finding 
symmetric matrices of order $n$ whose rows (or columns) sum $n$ given
numbers. That is, to solve a linear system of $n$ equations 
with $\frac 1 2 n(n+1)$ unknowns. 
For $n>1$ the solutions determine an affine 
space of dimension $\frac 1 2 n(n+1)-n=\frac 1 2 n(n-1)$. 
Moreover, for $n>1$ the second 
condition determine always an affine space of dimension $n-1$.
The well known case of of $n=1$ \cite{InfHul,CarRamdos} gives unique
solution to both conditions.
However, the constant $t$ remains always undetermined.

\bigskip

We will try to find now other generalizations of Shape Invariant potentials
which depend on $n$ parameters transformed by means of a translation. 
We should try a generalization using  inverse powers
of the parameters $m_i$; we know already that for the case $n=1$ 
there appear at least three new families of solutions 
(see Table~6 in \cite{CarRamdos}). 
So, we will try a solution of the following type, 
provided $m_i\neq 0$, for all $i$,
\begin{equation}
k(x,m)=\sum_{i=1}^n \frac{f_i(x)}{m_i}
+g_0(x)+\sum_{i=1}^n m_i g_i(x)\ . \label{minv1_np}
\end{equation} 
Here, $f_i(x)$, $g_i(x)$ and $g_0(x)$ are functions of $x$ to 
be determined. Substituting into (\ref{lhs_np}) we obtain, after a 
little algebra,
\ba
&&L(m)-L(m+1)=
-\sum_{i,j=1}^n \frac{f_i f_j(1+m_i+m_j)}{m_i(m_i+1)m_j(m_j+1)}
-2g_0\sum_{i=1}^n \frac{f_i}{m_i(m_i+1)}                \nonumber\\
&&-2\sum_{i,j=1}^n \frac{m_j g_jf_i}{m_i(m_i+1)}
+2\sum_{i,j=1}^n \frac{g_j f_i}{m_i+1}                  
+\sum_{i=1}^n \frac{2 m_i+1}{m_i(m_i+1)}\frac{df_i}{dx}+\dots\ , \nonumber
\ea
where the dots represents the right hand side of (\ref{lhs_npar}).
The coefficients of each of the different  dependences on the
parameters $m_i$  have to be constant. The term
$$
-\sum_{i,j=1}^n \frac{f_i f_j(1+m_i+m_j)}{m_i(m_i+1)m_j(m_j+1)}\,
$$
involves a symmetric expression under the interchange of the indices 
$i$ and $j$. As a consequence we obtain that $f_i f_j=\mbox{Const.}$ 
for all $i,j$. Since $i$ and $j$ run independently
the only possibility is that $f_i=\mbox{Const.}$ for all 
$i\in\{1,\,\dots,\,n\}$. We will assume that at least one of the 
$f_i$ is different from zero, otherwise we would be in the
already studied case. Then, the term
$$
-2g_0\sum_{i=1}^n \frac{f_i}{m_i(m_i+1)}\,,\quad\quad
$$
gives us $g_0=\mbox{Const.}$ and the term which contains
the derivatives of the $f_i$'s vanishes.
The sum of the terms
$$
2\sum_{i,j=1}^n \frac{g_j f_i}{m_i+1}
-2\sum_{i,j=1}^n \frac{m_j g_jf_i}{m_i(m_i+1)}
$$
is only zero for $n=1$. Then, for $n>1$
the first of them provides us $\sum_{i=1}^n g_i=\mbox{Const.}$ and 
the second one, $g_i=\mbox{Const.}$ for all $i\in\{1,\,\dots,\,n\}$. 
This is just a particular case of the trivial solution.
For $n=1$, however, we obtain more solutions; is the case 
already discussed in \cite{InfHul,CarRamdos}. It should 
be noted that, in general, 
$$
2\sum_{i,j=1}^n \frac{g_j f_i}{m_i+1}
-2\sum_{i,j=1}^n \frac{m_j g_jf_i}{m_i(m_i+1)}\neq
2\sum_{i,j=1}^n \frac{f_i g_j}{m_i+1}\left(1-\frac{m_j}{m_i}\right)\,,
$$ 
as one could be tempted to write if one does not take care.
Using the last equation as being true will lead to incorrect results.
As a conclusion we obtain that the trial solution $k(x,m)$  corresponding
to that of the case 
 $n=1$  admits no non--trivial generalization to solutions
of the type (\ref{minv1_np}).

It can be shown that if we propose further generalizations to greater degree
inverse powers of the parameters $m_i$, the only solution is also a 
trivial one. For example, if we try a solution of type
\begin{equation}
k(x,m)=\sum_{i,j=1}^n\frac{h_{ij}(x)}{m_i m_j}
+\sum_{i=1}^n \frac{f_i(x)}{m_i}
+g_0(x)+\sum_{i=1}^n m_i g_i(x)\,, \label{minv2_np}
\end{equation} 
where $h_{ij}(x)=h_{ji}(x)$,
the only possibility we will obtain is that all involved functions of 
$x$ have to be constant. 

\bigskip

Now we try to generalize (\ref{mlin_np}) to higher positive powers.
That is, we will try now a solution of type
\begin{equation}
k(x,m)=g_0(x)+\sum_{i=1}^n m_i g_i(x)+\sum_{i,j=1}^n m_i m_je_{ij}(x)\,. 
\label{mquad_np}
\end{equation} 
Substituting into (\ref{lhs_np}) we obtain, after several calculations,
\ba
&&L(m)-L(m+1)=4\sum_{i,j,k,l=1}^n m_i m_j m_k e_{ij} e_{kl}
+4\sum_{i,j,k,l=1}^n m_i e_{ij} m_k (e_{kl}+g_k)        \nonumber\\
&&+2\sum_{i,j=1}^n m_i m_j 
\left(\sum_{k,l=1}^n(e_{kl}+g_l)e_{ij}+\frac{de_{ij}}{dx}\right)
                                                        \nonumber\\
&&+4\sum_{i,j=1}^n m_i e_{ij} 
\left(\sum_{k,l=1}^n(e_{kl}+g_l)+g_0\right)             \nonumber\\
&&+2\sum_{i=1}^n m_i 
\left(g_i\sum_{j,k=1}^n (e_{jk}+g_j)
+\frac{d}{dx}\sum_{k=1}^n(e_{ik}+g_i)\right)            \nonumber\\
&&+\sum_{i,j=1}^n(e_{ij}+g_i)\left(\sum_{k,l=1}^n(e_{lk}+g_l)+2g_0\right)
+\frac{d}{dx}\left(\sum_{i,j=1}^n(e_{ij}+g_i)+2g_0\right)\,.
                                                        \label{lhs_quad_np}
\ea
As in previous cases, the coefficients of each different type of dependence
on the parameters $m_i$ have to be constant. Let us analyze the term of 
higher degree i.e. the first term on the right hand side of 
(\ref{lhs_quad_np}). Since it contains a completely symmetric 
sum in the parameters $m_i$, the dependence on the functions $e_{ij}$ 
should also be completely symmetric in the corresponding indices. For that
reason, we rewrite it as
$$
4\sum_{i,j,k,l=1}^n m_i m_j m_k e_{ij} e_{kl}=
\frac 4 3 \sum_{i,j,k,l=1}^n m_i m_j m_k (e_{ij} e_{kl}
+e_{jk} e_{il}+e_{ki} e_{jl})\,,
$$
from where it is found the necessary condition
$$
\sum_{l=1}^n(e_{ij} e_{kl}
+e_{jk} e_{il}+e_{ki} e_{jl})=d_{ijk}
\,,\quad\forall\,i,j,k\in\{1,\,\dots,\,n\}\,,
$$
where $d_{ijk}$ are completely symmetric in their three indices  constants.
The number of independent equations of this type is just the number
of independent components of a completely symmetric  in its three 
indices tensor, each one running from 1 to $n$. This number is 
$\frac 1 6 n(n+1)(n+2)$. 
The number of independent variables $e_{ij}$ is
$\frac{1}{2}n(n+1)$ from the symmetry on the two indices. 
Then, the number of unknowns minus the number of equations is
$$
\frac 1 2 n(n+1)-\frac 1 6 n(n+1)(n+2)
=-\frac 1 6 (n-1)n(n+1)\,.
$$ 
For $n=1$ the system has
the simple solution $e_{11}=\mbox{Const}$. For $n>1$ the system
is not compatible and has no solutions apart from the trivial one
$e_{ij}=\mbox{Const.}$ for all $i,j$. In either of these cases, 
it is very easy to deduce from the other terms in (\ref{lhs_quad_np})
that all of the remaining functions have to be constant as well,
provided that not all of the constants $e_{ij}$ vanish. 
For higher positive power dependence on the parameters $m_i$'s a similar result
holds. In fact, let us suppose that the higher order term in our 
trial solution is of degree $q$,
$\sum_{i_1,\,\dots,\,i_q=1}^n m_{i_1}m_{i_2}\cdots m_{i_q} 
T_{i_1,\,\dots,\,i_q}\,,$
where $T_{i_1,\,\dots,\,i_q}$ is a completely symmetric tensor in its
indices. Then, is easy to prove that the higher order term appearing
after substitution in (\ref{lhs_np}) 
is a sum whose general term is of degree $2q-1$
in the $m_i$'s. It is completely symmetric under interchange of their indices,
and appears the product of $T_{i_1,\,\dots,\,i_q}$ by itself but with one index 
summed. One then has to symmetrize the expression of the $T$'s in order
to obtain the independent equations to be satisfied, which is equal 
to the number of independent components of a completely symmetric tensor in its 
$2q-1$ indices. This number is $(n+2(q-1))!/(2q-1)!(n-1)!$. The number 
of independent unknowns is $(n+q-1)!/q!(n-1)!$. So, the number
of unknowns minus the one of equations is 
$$
\frac{(n+q-1)!}{q!(n-1)!}-\frac{(n+2(q-1))!}{(2q-1)!(n-1)!}\,.
$$
This number vanishes always for $n=1$, which means that the problem is
determined and we obtain that $T_{1,\,\dots,\,1}=\mbox{Const.}$, in agreement
with \cite[p. 28]{InfHul}. If $n>1$,
one can easily check that for $q>1$ that number is negative and hence
there cannot be other solution  apart from the trivial solution 
$T_{i_1,\,\dots,\,i_q}=\mbox{Const.}$ for all 
$i_1,\,\dots,\,i_q\in\{1,\,\dots,\,n\}$. From the terms of lower degree one
should conclude that the 
only possibility is a particular case of the trivial solution.

\section{Conclusions and outlook}

Let us comment on the relevance of the more important 
result of this paper, that
is, the fact that we have been able to solve
 the differential equation system  
(\ref{eqsgi}) and (\ref{eqsg0}). That problem was posed, but not solved, 
in a often cited paper by Cooper, Ginocchio and Khare in 
Physical Review {\bf D} \cite[pp. 2471--2]{CoopGinKha}. They use
a slightly different notation but it can be identified  their
formulas $(6.24)$  with our (\ref{mlin_np}) and 
our procedure by an appropriate redefinition of the 
parameters taking into account (\ref{parmi}) and (\ref{tas}). 
However, they failed to find any solution to these equations 
(for $n>2$), and believed that such a solution could hardly exist.


The conclusion is conceptually of great importance. 
That is,  
it has been made clear that an arbitrary but finite number
of parameters subject to transformation is not a limitation
to the existence of Shape Invariant partner potentials, 
and hence, to the existence of exactly solvable problems
in Quantum Mechanics. This leaves the door open to the 
possibility to pose and maybe solve further generalizations. 
We have as well the possibility of englobing particular cases
of known Shape Invariant partner potentials spread over the
extensive literature in the subject 
(see e.g. \cite{CoopKhaSuk} and references therein) into one
simple but powerful scheme of classification. In this sense,
we think the solution we have found here is very important
as it completes the excellent work started in 
\cite{CoopGinKha}.

Other conceptual point of great importance
is that we have gained much more generality in the solution to the
problem by a particularly simple but powerful idea. That is,
to consider the general solution of the Riccati equation with constant 
coefficients which gives all subsequent solutions, 
rather than particular ones. For doing this it has been of great
use the important properties of the Riccati equation.   

As a byproduct of our present results and that of 
\cite{CarRamdos} it is not difficult to see that for $n=1$ most 
of the solutions contained in \cite[Sec. {\bf VI}]{CoopGinKha},
later reproduced e.g. in \cite{CoopKhaSuk}, are directly 
related to some results of the classic paper \cite{InfHul}, since 
they are solutions of essentially the same equations.

\ack
One of the authors (A.R.) thanks the Spanish Ministerio de 
Educaci\'on y Cultura for a FPI grant, research project 
PB96--0717. Support of the Spanish DGES (PB96--0717) is also acknowledged.

\vfill
\eject

\begin{table}
\caption{General solutions of the equations  
(\ref{ric_y}) and (\ref{lin_z}). $A$,
$B$ and  $D$ are integration constants. The constant
$B$ selects the particular solution of (\ref{ric_y}) in each case.}
\label{sols_gens}
\begin{tabular*}{\textwidth}{@{}l*{15}{@{\extracolsep{0pt plus12pt}}l}}
\br
\multicolumn{1}{c}{\bt Sign of $a$\et}
        &\multicolumn{1}{c}{\bt$y(x)$\et}
                &\multicolumn{1}{c}{\bt $z(x)$\et}                   \\
\mr
                &               &                               \\    
\bt\quad$a=c^2>0$\et 
&\bt\quad$c\,\frac{B\,\sinh(c(x-A))-\cosh(c(x-A))}
{B\,\cosh(c(x-A))-\sinh(c(x-A))}$\et
  &\bt\quad\quad$\frac{\frac{b}{c}\{B\,\sinh(c(x-A))-\cosh(c(x-A))\}+D}
  {B\,\cosh(c(x-A))-\sinh(c(x-A))}$\et                          \\
                &               &                               \\
\mr
                &               &                               \\       
\bt\quad$a=0$\et 
                &\bt\quad\quad$\frac{B}{1+B(x-A)}$\et
    &\quad\quad\bt$\frac{b(\frac B 2 (x-A)^2+x-A)+D}{1+B(x-A)}$\et        \\
                &               &                               \\         
\mr
                &               &                               \\        
\bt\quad$a=-c^2<0$\et
  &\bt\quad$-c\,\frac{B\,\sin(c(x-A))+\cos(c(x-A))}
  {B\,\cos(c(x-A))-\sin(c(x-A))}$\et
 &\bt\quad\quad$\frac{\frac{b}{c}\{B\,\sin(c(x-A))+\cos(c(x-A))\}+D}
 {B\,\cos(c(x-A))-\sin(c(x-A))}$\et                             \\
                &               &                               \\
\br
\end{tabular*}
\end{table}

\begin{table}
\caption{
General solutions for the differential equation system
(\ref{ricgcm}), (\ref{linvj}) and (\ref{ling0}). 
$A$, $B$, $D_0$ and $D_j$ are arbitrary constants. The constant
$B$ selects the particular solution of (\ref{ricgcm}) 
for each sign of $nc_{cm}$.}
\label{ngcmvjg0}
\begin{tabular*}{\textwidth}{@{}l*{15}{@{\extracolsep{0pt plus12pt}}l}}
\br
\multicolumn{1}{c}{\bt Sign of $n c_{cm}$\et}
        &\multicolumn{1}{c}{\bt$n g_{cm}(x)$\et}
        &\multicolumn{1}{c}{\bt$v_j(x)$ for 
        $j\in\{2,\,\dots,\,n\}$ and  $g_0(x)$\et}               \\
\mr
                &               &                               \\    
\bt$n c_{cm}=C^2>0$\et  
&\bt\quad$ C f_{+}(x,A,B,C)$\et
 &\bt\quad$\frac{c_j-c_{cm}}{C}f_{+}(x,A,B,C)+D_j h_{+}(x,A,B,C)$\et    \\
                &               &                               \\
& & \bt\quad$\frac{c_0}{C}f_{+}(x,A,B,C)+D_0 h_{+}(x,A,B,C)$\et \\
                &               &                               \\
\mr
                &               &                               \\       
\bt$n c_{cm}=0$\et 
                &\bt\quad$B f_0(x,A,B)$\et
 &\bt\quad$(c_j-c_{cm})h_0(x,A,B)+D_j f_0(x,A,B)$\et                    \\
                &               &                               \\
         &  &\bt\quad$c_0 h_0(x,A,B)+D_0 f_0(x,A,B)$\et         \\
                &               &                               \\         
\mr
                &               &                               \\        
\bt$n c_{cm}=-C^2<0$\et 
&\bt\quad$ -C f_{-}(x,A,B,C)$\et
 &\bt\quad$\frac{c_j-c_{cm}}{C}f_{-}(x,A,B,C)+D_j h_{-}(x,A,B,C)$\et    \\
                &               &                               \\
& & \bt\quad$\frac{c_0}{C}f_{-}(x,A,B,C)+D_0 h_{-}(x,A,B,C)$\et \\
                &               &                               \\
\mr
\end{tabular*}
\begin{tabular*}{\textwidth}{@{}l*{15}{@{\extracolsep{0pt plus12pt}}l}}
\multicolumn{1}{c}{}
                        &\multicolumn{1}{c}{}                  \\
\bt where \et   &                                               \\
\bt\ \ $f_{+}(x,A,B,C)=\frac{B\,\sinh(C(x-A))-\cosh(C(x-A))}
{B\,\cosh(C(x-A))-\sinh(C(x-A))}$\et    
 &\bt\quad$h_{+}(x,A,B,C)
   =\frac 1 {B\,\cosh(C(x-A))-\sinh(C(x-A))}$\et                \\
                &                                               \\
\bt\ \ $f_{0}(x,A,B)=\frac{1}{1+B(x-A)}$\et     
 &\bt\quad$h_{0}(x,A,B)=\frac{\frac B 2 (x-A)^2+x-A}{1+B(x-A)}$\et      \\
                &                                               \\
\bt\ \ $f_{-}(x,A,B,C)=\frac{B\,\sin(C(x-A))+\cos(C(x-A))}
        {B\,\cos(C(x-A))-\sin(C(x-A))}$\et      
 &\bt\quad$h_{-}(x,A,B,C)=\frac{1}
        {B\,\cos(C(x-A))-\sin(C(x-A))}$\et                      \\      
                &                                               \\
\br
\end{tabular*}
\end{table}

\begin{table}
\caption{General solutions for $k(x,m)$ of the form (\ref{mlin_np}). 
$A$, $B$ are arbitrary constants. 
$\til D$ denotes the combination $\Dtil$, where $D_0$, $D_i$ are
the same as in Table~\ref{ngcmvjg0}.
The constant
$B$ selects the particular solution of (\ref{ricgcm}) 
for each sign of $nc_{cm}$.}
\label{sols_k_np_lin}
\begin{tabular*}{\textwidth}{@{}l*{15}{@{\extracolsep{0pt plus12pt}}l}}
\br
\multicolumn{1}{c}{\bt Sign of $n c_{cm}$\et}
 &\multicolumn{1}{c}{\bt$k(x,m)=g_0(x)+\sum_{i=1}^n m_i\,g_i(x)$\et} \\
\mr
                &                                               \\    
\bt\quad$n c_{cm}=C^2>0$\et     
&\bt\quad\quad$\frac 1 C (\sumc) f_{+}(x,A,B,C)
                +\til D h_{+}(x,A,B,C)$\et                      \\
                &                                               \\
\mr
                &                                               \\       
\bt\quad$n c_{cm}=0$\et 
 &\bt\quad\quad$(\sumc)h_0(x,A,B)+(\til D+B\summn) f_0(x,A,B)$\et       \\
                &                                               \\
\mr
                &                                               \\        
\bt\quad$n c_{cm}=-C^2<0$\et    
&\bt\quad\quad$\frac 1 C (\sumc) f_{-}(x,A,B,C)
                +\til D h_{-}(x,A,B,C)$\et                      \\
                &                                               \\      
\mr
\end{tabular*}
\begin{tabular*}{\textwidth}{@{}l*{15}{@{\extracolsep{0pt plus12pt}}l}}
\multicolumn{1}{c}{}                       			\\
\bt where \ $f_{+}=f_{+}(x,A,B,C)$,\quad $f_{0}=f_{0}(x,A,B)$,
\quad $f_{-}=f_{-}(x,A,B,C)$\et                                 \\      
\bt\qquad\quad$h_{+}=h_{+}(x,A,B,C)$,\quad $h_{0}=h_{0}(x,A,B)$,
\quad $h_{-}=h_{-}(x,A,B,C)$ 
        \quad are defined as in Table~\ref{ngcmvjg0}\et 	\\
                                                                \\
\br
\end{tabular*}
\end{table}

\begin{table}
\caption{Shape--Invariant partner potentials which depend on $n$
parameters transformed by traslation, when $k(x,m)$ is of the form
(\ref{mlin_np}) and $m=(m_1,\,\dots,\,m_n)$.
The Shape Invariance condition $\til V(x,m)=V(x,m-1)+R(m-1)$ 
is satisfied in each case. 
$A$, $B$ and $\til D$ are arbitrary constants.}
\label{sols_sipot_np}
\begin{tabular*}{\textwidth}{@{}l*{15}{@{\extracolsep{0pt plus12pt}}l}}
\br
\multicolumn{1}{c}{\bt Sign of $n c_{cm}$\et}
        &\multicolumn{1}{c}{\bt$V(x,m)-d$, $\til V(x,m)-d$ and  $R(m)$ 
when $k(x,m)=g_0(x)+\sum_{i=1}^n m_i g_i(x)$\et}\\
\mr
                &                                               \\    
\bt$n c_{cm}=C^2>0$\et  
&\bt$\frac{(\sumc)^2}{\suc} f_{+}^2
        +\frac{\til D}{C}(2(\sumc)+\suc) f_{+}h_{+}$\et         \\
        & \bt\quad \quad$+(\til D^2-(B^2-1)(\sumc))h_{+}^2$\et  \\
                &                                               \\
&\bt$\frac{(\sumc)^2}{\suc} f_{+}^2
        +\frac{\til D}{C}(2(\sumc)-\suc) f_{+}h_{+}$\et         \\
        &\bt\quad\quad$+(\til D^2+(B^2-1)(\sumc))h_{+}^2$\et    \\
                &                                               \\
                &\bt$R(m)=L(m)-L(m+1)=2(\sumc)+\suc$\et         \\
                &                                               \\
\mr
                &                                               \\
\bt$n c_{cm}=0$\et      
  &\bt$(\sumc)^2 h_0^2+(\til D+B\summn)
        (\til D+B(\summn+1))f_0^2$\et                           \\
&\bt\quad$+2 (\sumc)(\til D+B(\summn+\frac 1 2))f_0 h_0
-(\sumc)$\et                                                    \\
                &                                               \\
 &\bt$(\sumc)^2 h_0^2+(\til D+B\summn)
        (\til D+B(\summn-1))f_0^2$\et                           \\
&\bt\quad$+2 (\sumc)(\til D+B(\summn-\frac 1 2))f_0 h_0
+(\sumc)$\et                                                    \\
                &                                               \\
                &\bt$R(m)=L(m)-L(m+1)=2(\sumc)$\et              \\
                &                                               \\
\mr
                &                                               \\
\bt$n c_{cm}=-C^2<0$\et 
&\bt$-\frac{(\sumc)^2}{\suc} f_{-}^2
        +\frac{\til D}{C}(2(\sumc)+\suc) f_{-}h_{-}$\et         \\
&\bt\quad\quad$+(\til D^2-(B^2+1)(\sumc))h_{-}^2$\et            \\
                &                                               \\
&\bt$-\frac{(\sumc)^2}{\suc} f_{-}^2
        +\frac{\til D}{C}(2(\sumc)-\suc) f_{-}h_{-}$\et         \\
&\bt\quad\quad$+(\til D^2+(B^2+1)(\sumc))h_{-}^2$\et            \\
                &                                               \\
                &\bt$R(m)=L(m)-L(m+1)=2(\sumc)+\suc$\et         \\
                &                                               \\
\mr
\end{tabular*}
\begin{tabular*}{\textwidth}{@{}l*{15}{@{\extracolsep{0pt plus12pt}}l}}
\multicolumn{1}{c}{}                                           	\\
\bt where \ $f_{+}=f_{+}(x,A,B,C)$,\quad $f_{0}=f_{0}(x,A,B)$,
\quad $f_{-}=f_{-}(x,A,B,C)$\et                                 \\      
\bt\qquad\quad$h_{+}=h_{+}(x,A,B,C)$,\quad $h_{0}=h_{0}(x,A,B)$,
\quad $h_{-}=h_{-}(x,A,B,C)$ 
        \quad are defined as in Table~\ref{ngcmvjg0}\et 	\\
                                                                \\
\br
\end{tabular*}
\end{table}

\end{document}